\documentclass[nobibnotes,nofootinbib,aps,pre,showpacs,
preprint
]{revtex4-1}
\usepackage{setspace}
\usepackage{graphics}
\usepackage{graphicx}
\usepackage{color}
\usepackage{amsmath}
\usepackage{amssymb}
\usepackage{bm}
\usepackage{float}
\usepackage{epsfig}
\usepackage{epstopdf}
\usepackage{color,graphics}
\usepackage[toc,page]{appendix}

\begin{document}
\title{Collisional Effects, Ion-Acoustic Waves and Neutrino Oscillations}

\author{Fernando Haas and Kellen Alves Pascoal}
\affiliation{Instituto de F\'{\i}sica, Universidade Federal do Rio Grande do Sul, Av. Bento Gon\c{c}alves 9500, 91501-970 Porto Alegre, RS, Brasil}

\author{Jos\'e Tito Mendon\c{c}a}
\affiliation{IPFN, Instituto Superior T\'ecnico, Universidade de Lisboa, 1049-001 Lisboa, Portugal}
\affiliation{Instituto de F\'isica, Universidade de S\~ao Paulo, 05508-090 S\~ao Paulo, SP, Brasil}

\begin{abstract}
We analyze the role of collisional effects on the coupling between ion-acoustic waves and neutrino flavor oscillations, discussing its relevance for plasma instabilities in extreme plasma environments like in type II supernovae, where intense neutrino bursts exist. Electrons (leptons) are coupled to the electron-neutrino fluid through the weak Fermi force, but the electron-neutrinos are allowed to convert to other neutrino flavors and vice-versa. Due to the typically slow frequency of neutrino flavor oscillations, many orders smaller than e.g. the plasma frequency, an effective energy transfer between plasma waves and neutrino flavor oscillations take place at the low-frequency electrostatic branch, viz. the ion-acoustic mode. We show the destabilization of ion-acoustic waves in dense astrophysical scenarios, with a focus on the collisional effects mediated by electron-ion scattering. The maximal instability growth-rate is evaluated and compared to characteristic inverse times of type II supernova explosions. The results can be used for independent experimental verification of the non-zero neutrino mass, in a plasma physics context.  
\end{abstract}

\pacs{13.15.+g, 52.35.Pp, 97.60.Bw}
\maketitle

\section{Introduction}

In a recent work \cite{pre} the coupling between ion-acoustic waves (IAWs) and neutrino flavor oscillations was established using a model merging the well known models for neutrino-plasma interactions \cite{Bethe}-\cite{Silva} and for neutrino flavor oscillations 
\cite{Suekane}-\cite{Raffelt}. Due to the slow character of both IAWs and flavor oscillations, a powerful interaction between them takes place, with a corresponding plasma instability. Such instability is a suitable mechanism for supernova explosions, where abundant neutrino sources are recognized. The timeliness of the discovery is enhanced in view of the 2015's Physics Nobel Prize awarded to T. Kajita and A. B. McDonald by the empirical verification of neutrino flavor oscillations. Here a possible avenue for an alternative experimental test of these oscillations is also provided, in terms of collective plasma oscillations. Recently, also the impact of neutrino coupling on MHD waves was established, with a new perspective for joint astrophysical plasma and elementary particles approaches \cite{mhd}. The radial flux far from the center of the neutrinosphere can be accurately treated as a collimated beam, with small angular spread \cite{Silva}. Sources of anisotropic neutrino velocities distributions have been thoroughly analyzed \cite{Laming}.

The purpose of the present brief communication is to address the impact of collisional effects in the new instability due to the coupling of IAWs and flavor oscillations. For this purpose, the electrons and ions force equations of \cite{pre} will be modified by the addition of friction forces between the two plasma species, adapting the recipe for purely classical plasmas of Ref. \cite{Vranjes}. Here a fully ionized electron-ion plasma will be considered, viz. without a neutral component, which is the more likely scenario in dense astrophysical plasmas as in supernovae. In addition, the quasineutrality condition will be assumed, which is equivalent to the treatment in \cite{pre} where slow waves of frequency $\omega \ll \omega_{pi}$ where considered, where $\omega_{pi}$ is the ionic plasma frequency. As a warning to the reader, we remark that several steps involved in the calculation will be omitted, as long as they do not involve collisional effects. The complete treatment (but without collisions) is shown in detail in \cite{pre}, where frictional forces where not included in order to show the basic IAWs and flavor oscillations coupling in a more net way. However, the study of the collisional impact on such coupling is certainly necessary. More complex approaches for collisional IAWs could consider the Fokker-Planck collision operator \cite{Buti} instead of simple frictional forces.

The work is organized as follows. In Section II, the basic model is presented. In Section III, the procedure for the calculation of linear waves and instability growth rate is shown, together with explicit estimates in astrophysical settings. Section IV is reserved for the conclusions.

\section{Basic Model}

Denoting $n_{e,i}$ and ${\bf u}_{e,i}$ as respectively the electron (e) and ion (i) fluid densities and velocity fields, in a fluid setting one will have the continuity equations 
\begin{equation}
\frac{\partial n_e}{\partial t} + \nabla \cdot (n_e {\bf u}_e) = 0 \,,  \quad \frac{\partial n_i}{\partial t} + \nabla \cdot (n_i {\bf u}_i) = 0 \,, \label{eq01} 
\end{equation}
the electrons force equation
\begin{eqnarray}
m_e\left(\frac{\partial}{\partial t} + {\bf u}_e\cdot \nabla \right)\,{\bf u}_e = - \kappa_B T_e \,\frac{\nabla n_e}{n_e} &+& e\,\nabla\phi + \sqrt{2}\,G_F\,({\bf  E}_\nu + {\bf  u}_{e}\times{\bf  B}_\nu) \nonumber \\
&-& m_e \nu_{ei} ({\bf u}_e - {\bf u}_i) \,, \label{eq02}
\end{eqnarray}
and the ions force equation 
\begin{equation}
m_i\left(\frac{\partial}{\partial t} + {\bf u}_i\cdot \nabla \right)\,{\bf u}_i = - e\,\nabla\phi - m_i \nu_{ie} ({\bf u}_i - {\bf u}_e)\,, \label{eqi02} 
\end{equation}
where ions are assumed cold for simplicity. 
In Eqs. (\ref{eq02}) and (\ref{eqi02}), $m_{e,i}$ are the electron (charge $-e$) and ion (charge $+e$) masses, $\kappa_B$ is Boltzmann's constant, $T_e$ is the electron isothermal fluid temperature, $\phi$ is the electrostatic potential and $\nu_{ei}, \nu_{ie}$ are resp. the electron-ion and ion-electron collision frequencies. Due to global momentum conservation one has $m_e \nu_{ei} = m_i \nu_{ie}$. In addition, $G_F$ is Fermi's coupling constant, and ${\bf  E}_\nu, {\bf  B}_\nu$ are effective neutrino electric and magnetic fields,
\begin{equation}
 {\bf E}_\nu = - \nabla N_e - \frac{1}{c^2}\,\frac{\partial}{\partial t}\,(N_e {\bf  v}_e) \,, \quad {\bf B}_\nu = \frac{1}{c^2}\,\nabla  \times (N_e {\bf  v}_e) \,, \label{eq04}
\end{equation}
where $N_e, {\bf  v}_e$ are the electron-neutrino fluid density and velocity field and $c$ the speed of light. For completeness, we include Poisson's equation
\begin{equation}
\label{poi}
\nabla^2\phi = \frac{e}{\varepsilon_0}\,(n_e - n_i) \,,
\end{equation}
where $\varepsilon_0$ the vacuum permittivity constant. However, for simplicity the quasineutrality assumption will be assumed, so that Poisson's equation will be avoided for the most part.   

Denoting $N_\mu, {\bf  v}_\mu$ as the muon-neutrino fluid density and velocity field, one has two-flavor neutrino oscillations mediated by 
\begin{eqnarray}
 \frac{\partial N_e}{\partial t} + \nabla \cdot (N_e {\bf v}_e) &=& \frac{1}{2}\,N\,\Omega_0\, P_2 \,, \label{eq05} \\
 \frac{\partial N_\mu}{\partial t} + \nabla \cdot (N_\mu {\bf v}_\mu) &=& -\,\frac{1}{2}\,N\,\Omega_0\, P_2 \,, \label{eq06}
\end{eqnarray}
where $N = N_e + N_\mu$ is the total neutrino fluid number density and $P_2$ pertains to the quantum coherence contribution in a flavor polarization vector ${\bf  P} = (P_1, P_2, P_3)$. Moreover, 
$\Omega_0 = \omega_0 \sin 2\theta_0$, where $\omega_0 = \Delta m^2 c^4/(2\,\hbar\,{\cal E}_0)$ with $\Delta m^2$ being the squared neutrino mass difference. In addition, ${\cal E}_0$ is the neutrino spinor's energy in the fundamental state and $\theta_0$ is the neutrino oscillations mixing angle. The convective terms on the left-hand sides of Eqs. (\ref{eq05}) and (\ref{eq06}) are due to the neutrino flows, while the right-hand sides are neutrino sources due purely to flavor conversion. 

The neutrino force equations are  
\begin{eqnarray}
 \frac{\partial {\bf p}_e}{\partial t} + {\bf v}_e \cdot \nabla {\bf p}_e &=& \sqrt{2}\,G_F \left(
- \nabla n_e - \frac{1}{c^2}\,\frac{\partial}{\partial t}\,(n_e{\bf u}_e) + \frac{{\bf v}_e}{c^2} \times \left[\nabla\times(n_e{\bf u}_e)\right]
\right) \,, \label{eq07} 
\\
 \frac{\partial {\bf p}_\mu}{\partial t} + {\bf v}_\mu \cdot \nabla {\bf p}_\mu &=& 0 \,, \label{eq08}
\end{eqnarray}
where ${\bf p}_e = \mathcal{E}_e {\bf v}_e/c^2, \,{\bf p}_\mu = \mathcal{E}_\mu {\bf v}_\mu/c^2$ are the electron and muon neutrino relativistic momenta, and $\mathcal{E}_e, \mathcal{E}_\mu$ are the neutrino beam energies. 

The flavor polarization vector ${\bf P} = (P_1, P_2, P_3)$ equations in a material medium are given \cite{Suekane, Raffelt} by 
\begin{equation}
 \frac{\partial P_1}{\partial t} = -\Omega(n_e)P_2 \,, \quad 
 \frac{\partial P_2}{\partial t} = \Omega(n_e)P_1 - \Omega_0 P_3 \,, \quad 
 \frac{\partial P_3}{\partial t} = \Omega_0 P_2 \,, \label{eqx10}
\end{equation}
where $\Omega(n_e)= \omega_0 [\cos 2 \theta_0 - \sqrt{2}\,G_F\, n_e/(\hbar\omega_0)]$. In a given point of space, one has $\partial|{\bf P}|^2/\partial t = 0$. However, the fluctuations of the electron fluid density change the neutrino oscillations in space and time. For convenience we also define the neutrino-flavor oscillatory frequency $\Omega_\nu = \sqrt{\Omega^2(n_0)+\Omega_0^2}$, where $n_0$ is the equilibrium electron (and ion) number density.

The general model was thoroughly discussed in \cite{pre}, but with $\nu_{ei} = \nu_{ie} = 0$ therein. Here we perform the relevant steps including collisional effects. 

\section{Linear waves}

Consider the homogeneous static equilibrium for Eqs. (\ref{eq01})-(\ref{eqi02}), (\ref{poi})-(\ref{eq06}) and (\ref{eq07})-(\ref{eqx10}), given by 
\begin{eqnarray}
n_e &=& n_i = n_0 \,, \quad {\bf u}_e = {\bf u}_i = 0 \,, \quad \phi = 0 \,, \nonumber \\
N_e &=& N_{e0} \,, \quad N_\mu = N_{\mu 0} \,, \quad {\bf v}_e = {\bf v}_{\mu} = {\bf v}_0  \,, \label{homo}
\end{eqnarray}
together with the equilibrium flavor polarization vector components 
\begin{equation}
\label{peq}
P_1 = \frac{\Omega_0}{\Omega_\nu} \,, \quad P_2 = 0 \,, \quad P_3  = \frac{\Omega(n_0)}{\Omega_\nu} = \frac{N_{e0}-N_{\mu 0}}{N_0} \,,
\end{equation}
where $N_{e0}, N_{\mu 0}$ are the electron and muon equilibrium neutrino fluid densities. 

We suppose a linearization around the equilibrium for plane wave perturbations, with a $\delta$ standing for first-order quantities. For instance, $n_e = n_0 + \delta n_e \, \exp[i({\bf k}\cdot{\bf r} - \omega t)]$. In this way, the continuity equations (\ref{eq01}) give 
\begin{equation}
\label{ce}
\omega \delta n_e = n_0 {\bf k}\cdot\delta{\bf u}_e \,, \quad \omega \delta n_i = n_0 {\bf k}\cdot\delta{\bf u}_i \,.
\end{equation}
The electrons and the ions force equations (\ref{eq02}) an (\ref{eqi02}) give 
\begin{eqnarray}
m_e \omega \delta{\bf u}_e = \frac{\kappa_B T_e}{n_0}\,{\bf k} \delta n_e - e {\bf k} \delta\phi + \frac{\sqrt{2} G_F}{c^2} \left((c^2 {\bf k} - \omega {\bf v}_0) \delta N_e - \omega N_{e0} \delta{\bf v}_e\right) - i m_e \nu_{ei} (\delta{\bf u}_e - \delta{\bf u}_i) \nonumber \\ \label{e}
\end{eqnarray}
and 
\begin{equation}
m_i \omega \delta{\bf u}_i =  e {\bf k} \delta\phi - i m_i \nu_{ie} (\delta{\bf u}_i - \delta{\bf u}_e) \label{i}
\end{equation}
Taking the scalar product of both sides of Eqs. (\ref{e}) and (\ref{i}) with ${\bf k}$, summing the results to eliminate $\delta\phi$, using Eq. (\ref{ce}) and $m_i \gg m_e$, and applying the quasineutrality condition $\delta n_i \approx \delta n_e$ yields
\begin{equation}
(\omega^2 - c_s^2 k^2)\delta n_e = \frac{\sqrt{2} n_0 G_F}{m_i c^2} \left((c^2 k^2 - \omega {\bf k}\cdot{\bf v}_0) \delta N_e - \omega N_{e0} {\bf k}\cdot\delta{\bf v}_e\right) \,, \label{apa}
\end{equation}
where $c_s = \sqrt{\kappa_B T_e/m_i}$ is the IAW speed. 

Apparently from Eq. (\ref{apa}) the collisional effects have disappeared altogether. This is manifestly true if the neutrino effects are switched off, as already known for classical IAWs in the quasineutrality approximation \cite{Vranjes}. However, in the enlarged context some care is needed since the electrons-neutrino velocity ${\bf v}_e$ is influenced by the electron fluid, which is collisional. Therefore in principle the quantity $\delta{\bf v}_e$ in the right-hand side of Eq. (\ref{apa}) can have a collisional contribution.  

To proceed and evaluate such frictional effects on $\delta{\bf v}_e$, notice that the steps involved in the derivation of Eq. (A3) of \cite{pre} remain valid, so that 
\begin{eqnarray}
\delta{\bf v}_e = \frac{\sqrt{2}\,G_F}{\mathcal{E}_0\,(\omega - {\bf k}\cdot{\bf v}_0)}\,\Bigl[c^2\,{\bf k}\,\delta n_e &-& n_0\,\omega\,\delta{\bf u}_e - \left({\bf k}\cdot{\bf v}_0\,\delta n_e - \frac{n_0\,\omega}{c^2}\,{\bf v}_0\cdot\delta{\bf u}_e\right)\,{\bf v}_0 \nonumber \\
&-& n_0\,\left({\bf v}_0\cdot\delta{\bf u}_e\,{\bf k} - {\bf k}\cdot{\bf v}_0\,\delta{\bf u}_e\right)\Bigr] \,, \label{vvv}
\end{eqnarray}
where in the right-hand side it was approximated $\mathcal{E}_{e0} \approx \mathcal{E}_0$. Hence it is found that the $\sim \delta{\bf v}_e$ contribution in Eq. (\ref{apa}) is at least of order ${\cal O}(G_F^2)$. Due to the smallness of Fermi's constant we then just need the classical expression for $\delta{\bf u}_e$, which is found from Eq. (\ref{e}) with formally $G_F \equiv 0$ in it. Inserting $\delta{\bf u}_i$ from Eq. (\ref{i}) in this classical equation, we get 
\begin{eqnarray}
m_e \omega \delta{\bf u}_e = \frac{\kappa_B T_e}{n_0}\,{\bf k} \delta n_e - e {\bf k} \delta\phi - i \nu_{ie}\delta{\bf u}_e + i \,\frac{m_e \nu_{ei}}{\omega + i \nu_{ie}} \left(\frac{e {\bf k} \delta\phi}{m_i} + i \nu_{ie}\delta{\bf u}_e\right) \,.
\end{eqnarray}
We don't need to solve this last equation for $\delta{\bf u}_e$ but just remark that it implies that $\delta{\bf u}_e$ is parallel to ${\bf k}$ in the classical approximation. Hence from the electrons continuity equation it is found 
\begin{equation}
\label{sem}
\delta{\bf u}_e \approx \frac{\omega {\bf k} \delta n_e}{n_0 k^2} \,,
\end{equation}
which inserted into Eq. (\ref{vvv}) gives 
\begin{equation}
\delta{\bf v}_e = \frac{\sqrt{2}\,G_F}{\mathcal{E}_0\,(\omega - {\bf k}\cdot{\bf v}_0)}\left(c^2 - \frac{\omega^2}{k^2}\right) \left({\bf k} - {\bf k}\cdot{\bf v}_0 \frac{{\bf v}_0}{c^2}\right) \delta n_e \,. \label{ve}
\end{equation}
The important point is that collisional effects have indeed disappeared altogether. Inserting Eq. (\ref{ve}) into Eq. (\ref{apa}), one regains Eq. (26) of \cite{pre}, reproduced here for convenience, 
\begin{eqnarray}
\Bigl(\omega^2 - c_s^2 k^2 + \frac{2\,G_{F}^2\,N_{e0}\,n_0\,\omega\,\left(c^2 k^2 - ({\bf k}\cdot{\bf v}_0)^2\right)}{m_i\,c^2\,\mathcal{E}_{0}\,(\omega - {\bf k}\cdot{\bf v}_0)}\Bigr)\,\delta n_e = \frac{\sqrt{2}\,G_F n_0}{m_i\,c^2}(c^2 k^2 - \omega\,{\bf k}\cdot{\bf v}_0)\,\delta N_e \,. \label{iaw}
\end{eqnarray}

A critical examination of the remaining steps detailed in \cite{pre} shows that these are not changed by collisional effects. Hence one still finds 
\begin{eqnarray}
 (\omega &-& {\bf k} \cdot {\bf v}_{0}) \,\delta N_e = \frac{\sqrt{2}\,G_F\, N_{e0}}{\mathcal{E}_{0}\,(\omega - {\bf k} \cdot {\bf v}_{0})} 
 \left(c^2\,k^2 - ({\bf k} \cdot {\bf v}_{0})^2\right) \delta n_e + \frac{\sqrt{2}}{2} \,\frac{N_0\, \Omega_0^2 \omega G_F\,\delta n_e}{(\omega^2 - \Omega_\nu^2)\hbar\Omega_\nu} \label{eqo} \,.  
\end{eqnarray}
%
Here $\mathcal{E}_0 = \mathcal{E}_{e0} \approx \mathcal{E}_{\mu 0}$ is the equilibrium quasi-mono-energetic neutrino beam energy.

From Eqs. (\ref{iaw}) and (\ref{eqo}) one then find the dispersion relation 
\begin{eqnarray}
\omega^2 = c_s^2\,k^2 + \frac{\Delta_e\,c^2\,k^2\,\Lambda(\theta)\,(c^2\,k^2 - \omega^2)}{(\omega-{\bf k}\cdot{\bf v}_0)^2} + \frac{\Delta\,\Omega_0^2\,\omega\,\mathcal{E}_0\,(c^2\,k^2 - \omega\,{\bf k}\cdot{\bf v}_0)}{2\,\hbar\,\Omega_\nu\,(\omega - {\bf k}\cdot{\bf v}_0)\,(\omega^2 - \Omega_{\nu}^2)} \,, \label{disp}
\end{eqnarray}
where  
\begin{equation}
\Delta_e = \frac{2\,G_F^2\,N_{e0}\,n_0}{m_i\,c^2\,\mathcal{E}_0} \,, \quad \Delta = \frac{2\,G_F^2\,N_0\,n_0}{m_i\,c^2\,\mathcal{E}_0} \,, \quad \Lambda(\theta) = \left(1 - \frac{v_0^2}{c^2}\right)\,\cos^{2}\theta + \sin^{2}\theta \,,
\end{equation}
with ${\bf k}\cdot{\bf v}_0 = k\,v_0\,\cos\theta$. 

Formally setting $\Delta = 0$ in Eq. (\ref{disp}) would reproduce Eq. (13) of \cite{Monteiro}, which describes neutrino-plasma IAWs without flavor conversion, taking into account $c_s \ll c$ which is necessary for non-relativistic electrons. From inspection, one will have a strong coupling between IAWs, the neutrino beam and the neutrino oscillations if and only if 
\begin{equation}
\label{drc}
\omega \approx c_s\,k = \Omega_\nu = {\bf k}\cdot{\bf v}_0 \,. 
\end{equation}
The corresponding growth-rate instability $\gamma$ was found in \cite{pre}, reading 
\begin{equation}
\label{gam}
\gamma = (\gamma_\nu^3 + \gamma_{\rm osc}^3)^{1/3} \,,
\end{equation}
where
\begin{equation}
\gamma_\nu = \frac{\sqrt{3}}{2}\,\left(\frac{\Delta_e}{2}\,\left(\frac{c}{c_s}\right)^4\right)^{1/3}\Omega_{\nu} \,, \quad \gamma_{\rm osc} = \frac{\sqrt{3}}{2}\,\left(\frac{G_{F}^2\,N_0\,n_0\,\Omega_{0}^2}{4\,\hbar\,\kappa_B\,T_e}\right)^{1/3} \,. \label{gs}
\end{equation}
The quantity $\gamma_\nu$ is due to the usual neutrino-plasma coupling, while the flavor conversion effect is entirely contained in $\gamma_{\rm osc}$.

\begin{figure}[!hbt]
\begin{center}
\includegraphics[width=8.0cm,height=6.0cm]{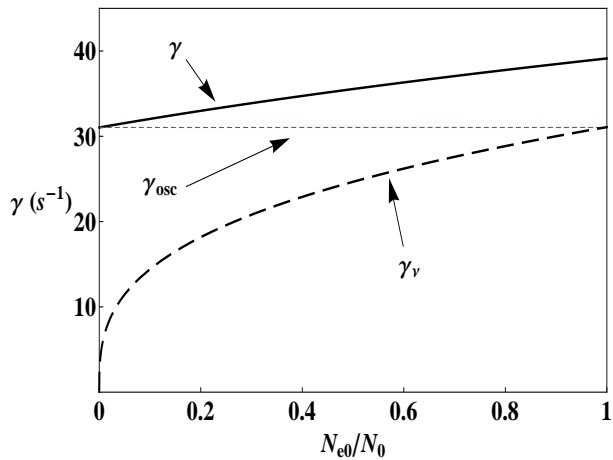}
\end{center}
\caption{Continuous line: growth rate $\gamma$ from Eq. (\ref{gam}) as a function of the normalized electron-neutrino population. Line-dashed curve: the growth rate $\gamma_\nu$ without taking into account neutrino oscillations. Horizontal dot-dashed line: the growth rate 
$\gamma_{\rm osc}$ associated to neutrino oscillations, from Eq. (\ref{gs}). Parameters: ${\cal E}_0 = 10 \,{\rm MeV}, N_0 = 10^{41}\,{\rm m}^{-3}, n_0 = 10^{35} \, {\rm m}^{-3}$, $\kappa_B T_e = 35 \,{\rm keV}$.}
\label{figure}
\end{figure}

As an example, we compute the growth rate in type II core-collapse supernovae settings, as in the supernova SN1987A with a neutrino flow of $10^{58}$ neutrinos of all flavors and energy between $10-15$ MeV \cite{Hirata}. We use $G_F = 1.45 \times 10^{-62} \, {\rm J.\,m^3}$ and take $\Delta m^2\,c^4 = 3 \times 10^{-5} \,({\rm eV})^2 \,,\sin(2\theta_0) = 10^{-1}$, which are appropriate to solve the solar neutrino problem \cite{Raffelt}. Moreover set ${\cal E}_0 = 10 \,{\rm MeV}, \kappa_B T_e = 35 \,{\rm keV}, N_0 = 10^{41}\,{\rm m}^{-3}, n_0 = 10^{35} \, {\rm m}^{-3}$. For these values $\omega \approx \Omega_\nu = 1.94 \times 10^7 \,{\rm rad}/s$, much smaller than $\omega_{pi} = 4.16 \times 10^{17}\,{\rm rad}/s$. In addition, $c_s = 1.83 \times 10^6 \, {\rm m/s} \ll c$ and $k = \Omega_{\nu}/c_s = 10.61 \,{\rm m}^{-1}$, corresponding to a wavelength $\lambda = 2\pi/k = 0.59 \, {\rm m}$. Finally, $\gamma_{\rm osc} = 31.03 \, {\rm s}^{-1}$ and the maximal growth rate is $\gamma_{\rm max} = 39.13 \, {\rm s}^{-1}$. Therefore,  $1/\gamma_{\rm max} \sim 0.03 \, {\rm s}$, much smaller than the accepted characteristic time of supernova explosions  around 1 second. Figure \ref{figure} shows the numerical value of instability growth rate, as a function of the initial normalized electron-neutrino population.  The difference in comparison to the estimates in \cite{pre} is that here a smaller plasma temperature enhances $\gamma_\nu$ in comparison to $\gamma_{\rm osc}$ so that for a pure electron-neutrino initial condition ($N_{e0} = N_0$) they are equal. Otherwise, $\gamma_{\rm osc} > \gamma_\nu$, showing the dominant role of neutrino flavor oscillations in this case. 

\section{Conclusion}

In this work we have shown that the coupling between IAWs and neutrino flavor oscillations in a completely ionized plasma is not affected by collisional effects, confirming the instability growth-rate estimates described in \cite{pre}. For simplicity, the quasineutrality condition was assumed. An intuitive explanation for the result is that in an ideal (infinite conductivity) plasma the electron and ion fluid velocities adjust themselves so that $n_0 {\bf k}\cdot({\bf u}_e - {\bf u}_i) =  \omega (\delta n_e - \delta n_i) \approx 0$, so that the two-fluid friction forces become irrelevant. Such feature happens for very slow waves so that $\omega \ll \omega_{pi}$. A less immediate finding is that the collisional force on electrons gives just a higher-order contribution on the neutrino fluid momentum, as manifest in Eq. (\ref{ve}).  A complete treatment beyond quasineutrality would deserve the full account of Poisson's equation, which although certainly feasible is much more cumbersome in view of the coupling with all the neutrino quantities. However, since the present analysis is devoted more to IAWs than to ionic waves, the quasineutrality approximation is well justified. Further improvements would involve a non-zero ionic temperature, magnetic fields, and degenerate and relativistic electrons.

\acknowledgments
F.~H.~ and J.~T.~M.~ acknowledge the support by Con\-se\-lho Na\-cio\-nal de De\-sen\-vol\-vi\-men\-to Cien\-t\'{\i}\-fi\-co e Tec\-no\-l\'o\-gi\-co (CNPq) and EU-FP7 IRSES Programme (grant 612506 QUANTUM PLASMAS FP7-PEOPLE-2013-IRSES), and K.~A.~P.~ack\-now\-ledges the support by Coordena\c{c}\~ao de Aperfei\c{c}oamento de Pessoal de N\'{\i}vel Superior (CAPES).

\end{document}